
\input harvmac

\rightline{FTUAM-94/2}
\bigskip

\centerline{ \bf A Renormalization Group Analysis}
\centerline{\bf of the NCG constraints $m_{top} = 2\, m_W$, $ m_{Higgs} =
3.14 \,m_W$}
\bigskip

\centerline{ \bf E. \'Alvarez, J.M. Gracia-Bond\'{\i}a \foot{And
Departamento de F\'{\i}sica Te\'orica, Universidad de Zaragoza, 50009
Zaragoza, Spain}
and C.P. Mart\'{\i}n}
\bigskip
\centerline{Departamento de F\'{\i}sica Te\'orica,
CXI}\centerline{Universidad Aut\'onoma}
\centerline{28049 Madrid, Spain}

\vskip .3in
We study the evolution under the renormalization group of the
restrictions on the parameters of the standard model coming from
Non-Commutative Geometry, namely $ m_{top} = 2\, m_W$ and $ m_{Higgs} =
3.14\, m_W$. We adopt the point of view that these
relations are to be interpreted as {\it tree level} constraints, and,
as such, can be implemented in a mass independent renormalization
scheme only at a given energy scale $\mu_0$. We show that
the physical predictions on the top and Higgs masses  depend weakly on $\mu_0$.

\Date{1/94}

\newsec{Non-Conmutative Geometry constraints on the parameters of the
Standard Model}
There has been recent interest in the field theoretical
\ref\frohlich{A.H. Chamseddine,G. Felder and J. Fr\"ohlich
Nucl. Phys. B395 (1993) 672.}  applications of Connes'
generalization of ordinary geometry
     \ref\connes{A. Connes, Pub. Math. IHES, 62, (1985), 257}, termed by him
Non-Commutative geometry (NCG fron now on). Connes' approach amounts to a new
 gauge principle incorporating the Higgs sector together with the pure
gauge terms. By the same token, the Dirac operator incorporates the matrix
 of Yukawa
 couplings between the chiral fermions and the Higgs field. Apparently,
 the Connes-Lott
derivation of the standard model of the strong and electroweak interactions
yields fewer arbitrary parameters than  the Glashow-Weinberg-Salam
 formulation of it.  There is some argument about this point
\ref\coque{R.Coquereaux,J. Geom. Phys. 11,(1993),307} but in the last
 version   {\ref\conlast{A. Connes, ``Non-Commutative Geometry'', IHES preprint
 (1993).}, Connes claims that there are two, and only two, such relationships,
 namely:

\eqn\top{m_{top} = 2 \,m_W}
\eqn\higgs{m_{higgs} = 3.14 \,m_W}

Note that the quotient of these relations can be obtained independently by
 considering the fermionic part of the lagrangian in NCG
\ref\vgb{J.C.Varilly and J.M.Gracia-Bond\'{\i}a,J.Geom.
 Phys. 12 (1993) 223}.
\bigskip
Our point of view  (lacking a more fundamental option) is
that once obtained, the NCG lagrangian is an ordinary quantum field
theoretical one, in which some constraints are imposed at the tree level.
Using standard techniques (\ref\collins{J.C.Collins,"Renormalization",
(Cambridge University Press)}, \ref\zimm{W.Zimmermann,Comm.Math.Phys.,
97,(1985),142})
we showed analytically in a recent letter
\ref\agm{E.\'Alvarez,J.M.Gracia-Bond\'{\i}a and C.P. Mart\'{\i}n,Phys.Lett.
B306 (1993) 55 }
that (at least in a toy model) those constraints could not be implemented
in a renormalization-group invariant way; physically this means that if we
impose them at one scale, say $\mu_0$, then the RG evolution would violate
them.

Our aim in the present letter is to address this same issue for the
standard model itself (we shall find the same result as we did before in a
simplified situation) as well as to draw some quantitative estimates on the
amount of evolution of the relations themselves, as compared to some
physically relevant quantity.
\bigskip
We shall present a certain amount of evidence that this evolution is
``small'' (in a well-defined sense), but we shall not venture
any further hypothesis on the physical origin of this fact.

\newsec{The evolution of the NCG constraints under the renormalization group
in the one-loop approximation}
In spite of the non-existence of a fully self-consistent and systematic set
of beta-functions (in particular, with respect to the treatment of the
$\gamma_5$ \`a la 't Hooft and Veltman), we have borrowed the expressions of
the one-loop beta functions (in the $\overline{ MS}$
 scheme) from the literature:
\ref\macha{M.E. Machacek and M.T. Vaughin, Nucl. Phys. B222 (1983)
83;B236 (1984) 221; B249 (1985) 70}. Neglecting
all leptonic Yukawa couplings, and all the quark Yukawas but the top,
the corresponding beta functions are:
\eqn\betat{ 4 \pi \beta_t = \alpha_t ( 9 \alpha_t - 16 \alpha_3 - {9
\over 2} \alpha_2 -{17\over 10} \alpha_1)}
\eqn\betatr{4 \pi \beta_3 = - 14 \alpha_{3}^{2}}
\eqn\betado{ 4 \pi \beta_2 = - {19\over 3} \alpha_{2}^{2}}
\eqn\betaun{ 4 \pi \beta_1 = {41\over 5} \alpha_{1}^{2}}
\eqn\betah{ 4 \pi \beta_h = 6 \alpha_{h}^2 + 12 \alpha_h \alpha_t -
{9\over 5} \alpha_h \alpha_1 - 9 \alpha_h \alpha_2 + {27\over 50}
\alpha_{1}^2 + {9\over 5} \alpha_1 \alpha_2 + {9 \over 2} \alpha_{2}^2
- 24 \alpha_{t}^2}

Where $\beta_h$ is the beta function of the quartic Higgs coupling
constant, $ \alpha_h = {\lambda\over 4\pi}$; $\beta_t$ the corresponding
one for the top Yukawa coupling, $\alpha_t = {g_{t}^2\over 4\pi}$, and
$\beta_i(i=1,2,3)$ are the beta functions corresponding to the gauge coupling
constants of U(1) and U(2)
(with an SU(5) normalization) and SU(3). Our Weinberg angle is defined
through
$\alpha = \alpha_{2} sin ^2\theta_w = {3\over 5} \alpha_{1} cos^2\theta_w$.
The physical masses are given in terms of the vacuum expectation value
of the neutral component of the Higgs field, $v$,
$m_W = {g_2 v\over 2}$; $ m_{H}^2 = { \lambda v^2\over 2}$; $m_t =
{g_t v\over \sqrt{2}}$ (neglecting the Kobayashi-Maskawa phases).

The relations of NCG boil down  in
our notations to
\eqn\relun{\alpha_t - 2\, \alpha_2 = 0}
\eqn\reldo{\alpha_h - 4.93\, \alpha_2 = 0}
which are physically equivalent to predictions on the masses of the
top and Higgs particles:
\eqn\mt{ m_t = 2\, m_W}
\eqn\mh{m_H = 3.14\, m_W}
As we have already said, from our point of view, those are tree level
constraints. If we want to implement them in a mass-independent
renormalization scheme, we have to choose a particular scale $\mu_0$ at which
we assume them to be valid, and study afterwards the RG flow from
these  initial conditions.

To be specific, what we did was the following: we chose initial
conditions at a point $\mu_0$ corresponding
to(\ref\langacker{P.Langacker, ``Precision tests of the Standard
Model, Philadelphia preprint, UPR-0555T (1993)})
\eqn\initial{\alpha_3 (\mu= m_Z) = 0.12}
\eqn\initia{\alpha_2 (\mu = m_Z) = 0.034}
\eqn\initi{\alpha_1 (\mu = m_Z) = 0.017}
(in our approximation, these coupling constants form a closed set
under the renormalization group)
and, furthermore, we got the other two initial conditions from
imposing the Connes-Lott constraints at $\mu_0$:

\eqn\r{\alpha_t(\mu_0) =  2 \,\alpha_2(\mu_0)}
\eqn\s{\alpha_h(\mu_0) =  4.93\, \alpha_2(\mu_0)}
\bigskip

We have displayed in the table  the dependence on the physical
predictions on the scale $\mu_0$ at which  one chooses to impose the
Connes-Lott relations as initial conditions. As is apparent from the table,
this dependence never gets much bigger than 10\% , even if one is willing to
extrapolate the standard model as such as far as $10^8$ GeV, which is
the approximate location of the Landau pole for $\alpha_h$ with our
set of initial conditions .
\bigskip
$$\displaylines{
\matrix{\mu_0&m_{top}&m_{higgs}\cr
10&138&283\cr
92&160&252\cr
 200&166&244\cr
300&169&241\cr
400&171&239\cr
500&172&237\cr
600&173&236\cr
700&174&235\cr
800&175&234\cr
900&176&233\cr
10^3&176&233\cr
10^4&186&223\cr
10^5&192&217\cr
10^6&197&215\cr
10^7&199&213}}$$
\bigskip
\bigskip\bigskip\bigskip
It is sometimes held the viewpoint \ref\kastler{D. Kastler and T.
Sch\"ucker, Theor. Math. Phys. 92 (1992) 522}
that there are actually four
relationships instead of only two, depending besides on an unknown
parameter
$x$, to wit:
$$\alpha_3 = {1\over 2} ({4 - 2 x\over 1- x})^{1/2} \alpha_2$$
$$sin^2 \theta_W ={3(1 - x/2)\over 8 - 2x}$$
$$m_W = {1\over 2}({1-x\over 1 -x/2})^{1/2} m_t$$
$$m_H = (3 - {3\over 2} { 3 x^2 - 8 x + 5 \over 5 x^2 - 17 x +
14})^{1/2} m_t$$
We just would like to point out here that these relationships are
inconsistent. From the (known) evolution of $\alpha_2$ and $\alpha_3$
(we remind the reader that the initial conditions for the closed
subset $\alpha_1$, $\alpha_2$ and $\alpha_3$ are completely fixed
by the experiment, and we can not change them)
one determines the unique point (for each $x$) at which the first
relationship can consistently be imposed.
 For $x=0$ this gives $\mu_0 = 2\times 10^{17} GeV$.\foot{The only
other value of $x$ for which there is a solution is $x \sim 1$, for
which $\mu_0 \sim 10\, GeV$, and which predicts huge values for the
top and Higgs masses}

We should now impose all other relations at this value of $\mu_0$ and
get a {\it unique} prediction for $m_t$ and $m_H$, but there are two
problems: first of all, $\mu_0$ is well above the Landau ghost for
the Higgs self-coupling, so that one can not impose consistently
initial conditions on it. Besides, even if this were not the case, the
value one predicts for Weinberg's angle is incompatible with the
experiment. This is actually the main reason why we have stuck all
along the paper to the original Connes' interpretation of considering
two constraints only.

\newsec{Conclusions}

Is all this physically significant?

Perhaps, at least if the top mass turns out to be experimentally close to
160 GeV.
The figures are small in absolute value (we are always talking of
dimensionless quantities). Still, does this mean anything physically?

Let us take the conservative point of view that if we impose a random
constraint on the physical coumplings, say,
\eqn\random{\lambda \equiv \sum c_i \alpha_i = 0}
at a given scale, $\mu_0$, the natural expectation is to find that at another
scale, $\mu$,  $\lambda$ will not be zero, but instead, it will be of the
order of
\eqn\alambda{\lambda (\mu) \sim max_{i} [c_i (\alpha_i(\mu) -
\alpha_i(\mu_o))]}
We would be tempted to say that when the coresponding
quotient \eqn\q{q \equiv {\lambda (\mu) \over max_{i} [c_i (\alpha_i(\mu) -
\alpha_i(\mu_o))]}}
is smaller than one, the relation between coupling constants is "better than
random" (whatever explanation it has,

 it is obvious that it is an
statistical fact that there are roughly equal numbers of "good" than "bad"
relations in the "space" of all possible relations; still, one would be
inclined to look for a physical explanation only if this quantity is
appreciably smaller than one).

In our case $q$ turns ot to be
\eqn\stop{q(top) = 0.7}
\eqn\shiggs{q(higgs) = 0.4}
if $\mu_0$ ranges from $m_Z$ up to $1$ TeV.

A further point is the following: in a celebrated paper, Veltman
\ref\tini{M. Veltman, Acta Phys. Pol.B12 (1981) 437 \semi {J. Kubo,K.
Sibold and W. Zimmermann, Phys. Lett B220 (1989) 191}}
worked out the relations one
needed to have among the parameters of the standard model if one
wanted to have cancellation of quadratic divergences.
We have explicitely checked that Connes relations are not compatible
with Veltman's for any value of the parameter $x$ introduced in ref. \kastler .

{\bf Acknowledgements}
We are grateful to B. Gavela, M.J. Herrero, L. Ib\'a\~nez, and C. Mu\~noz,
for
enlightening discussions and/or reading the manuscript.
This work originated from a question of S. Weinberg to one of us (JMGB).
We  have been supported in part by CICYT (Spain).

\listrefs

\bye